\documentclass[conference]{IEEEtran}
\IEEEoverridecommandlockouts

\usepackage{subcaption}

\usepackage[ruled,linesnumbered]{algorithm2e}
\usepackage{amsmath,amssymb,amsfonts}
\usepackage{graphicx}
\usepackage{soul,color}
\usepackage{authblk}
\usepackage{float}
\usepackage{subcaption} 
\usepackage{titlesec}
\usepackage{lettrine} 
\usepackage{hyperref}
\usepackage{booktabs}
\DeclareUnicodeCharacter{FF0C}{,}

\setstcolor{red}

\usepackage[table]{xcolor} 

\begin{document}

\title{Transfer Learning for VLC-based indoor Localization: Addressing Environmental Variability}


\author[1,2]{Masood Jan}
\author[1]{Wafa Njima}
\author[1,3]{Xun Zhang}
\author[4]{Alexander Artemenko}

\affil[1]{\textit{Institut Sup\'{e}rieur d'Electronique de Paris (ISEP), Paris, France}}
\affil[2]{\textit{Sorbonne University, Paris, France}}
\affil[3]{\textit{University Paris-Saclay， LISV， France}}
\affil[4]{\textit{Robert Bosch GmbH, Stuttgart, Germany}

\authorcr Emails: \{masood.jan, wafa.njima, xun.zhang\}@isep.fr, alexander.artemenko@de.bosch.com}



\maketitle


\begin{abstract}

Accurate indoor localization is crucial in industrial environments. 
Visible Light Communication (VLC) has emerged as a promising solution, offering high accuracy, energy efficiency, and minimal electromagnetic interference. However, VLC-based indoor localization faces challenges due to environmental variability, such as lighting fluctuations and obstacles. To address these challenges, we propose a Transfer Learning (TL)-based approach for VLC-based indoor localization. Using real-world data collected at a BOSCH factory, the TL framework integrates a deep neural network (DNN) to improve localization accuracy by 47\%, reduce energy consumption by 32\%, and decrease computational time by 40\% compared to the conventional models. The proposed solution is highly adaptable under varying environmental conditions and achieves similar accuracy with only 30\% of the dataset, making it a cost-efficient and scalable option for industrial applications in Industry 4.0.

\end{abstract}


\section{Introduction}
\label{sec:intro}

\IEEEPARstart{I}{ndoor} localization is becoming increasingly crucial in modern industrial environments. With increasing automation and smart systems, accurate localization is essential for ensuring operational efficiency, safety, and productivity \cite{hayward2022survey, umer2020use}.
Traditional localization systems 
are often limited in indoor environments due to signal interference and low accuracy in areas with dense structures and various obstacles \cite{leitch2023indoor}. This makes indoor localization systems particularly challenging, especially in large-scale industrial settings. 

Recently, Visible Light Communication (VLC) has emerged as a promising technology for indoor localization, 
especially where radio-frequency signals suffer from interference or congestion.
VLC provides high data rates, minimal electromagnetic interference, and excellent accuracy in indoor positioning systems \cite{wang2024survey}. Recent studies have explored novel techniques, including deep learning models, adaptive signal processing, and hybrid VLC-RF systems, to further enhance localization accuracy and robustness \cite{zhu2024survey}. 
These advancements help address challenges like environmental noise, signal degradation, and multi-path interference, further strengthening VLC's potential for industrial localization. Moreover, VLC is energy-efficient, using Light Emitting Diode (LED) lights for both illumination and communication, making it cost-effective and scalable \cite{almadani2020visible, nguyen2024mobile}.
However, despite its promising advantages, VLC-based indoor localization face a key challenge: environmental variability. The performance of such systems is highly sensitive to changes in lighting conditions caused by daylight fluctuations, artificial lighting adjustments and obstacles. These variations introduce noise into the received signals, reducing the model's ability to generalize across different scenarios and time periods effectively, thus leading to high localization errors, reduced success rates, and increased energy consumption due to longer training times as models struggle to adapt to these noisy, changing environments 
\cite{lin2018indoor}. 

To overcome this challenge, machine learning (ML)-based localization models have been used. However, these models typically assume that the trained model remains fixed over time, space, or devices, enabling its application in an online setting without further adaptation. In industrial environments, this assumption is often not valid due to  changes in data distribution over time and spatial variability across large industrial spaces \cite{pan2008transfer}. As a result, periodic model updates are essential, however, collecting large datasets in such environments can be impractical due to the cost of data acquisition. 
To deal with these challenges, innovative methods like Federated Learning (FL) have been explored which enables decentralized model training across devices \cite{jan2024privacy}. 
However, it requires frequent updates to handle dynamic environments. 
Continuous Learning (CL) presents a solution by enabling models to adapt incrementally to new data over time \cite{wickramasinghe2023continual}. However, it suffers from catastrophic forgetting, where previously learned information is overwritten by new data. 
In this context, Transfer Learning (TL) emerges as a powerful approach to address CL's limitations. It leverages the knowledge gained from a source domain to improve learning performance or reduce data requirements in a target domain. This makes TL suited for addressing environmental variability. 
It has demonstrated its versatility across a wide array of applications, including image classification, sentiment analysis, dialog systems, text mining, \cite{long2015fully, raina2007self} and natural language processing \cite{wei2016transfer}. 



Recent studies have proposed various approaches to improve indoor localization systems using VLC in industrial environments. In \cite{almadani2020visible}, a 3D Visible Light Positioning (VLP) system was explored for real-time tracking in dynamic environments, showing accuracy despite varying signal strengths. However, it did not account for environmental factors 
as fluctuating lighting conditions and multi-path interference, 
significantly affect the reliability of VLC in large-scale deployments. \cite{zhu2024survey} presents a localization system based on VLC, demonstrating its ability for real-time positioning in industrial settings. However, this study overlooked environmental noise and signal degradation, highlighting an opportunity to address these issues in complex industrial setups where line-of-sight is often obstructed.

The overall contributions of this paper can be summarized as follows:

\begin{itemize}

\item \textbf{Real world VLC data collection and pre-processing:} Our dataset is particularly novel as it captures the complexities of an operational industrial environment. Collected from a production line at the BOSCH factory in Blaichach, Germany, it reflects real-world signal variations and interference patterns. Unlike many existing studies that rely on controlled or simulated datasets, our dataset provides a realistic and challenging benchmark for VLC-based indoor localization.

\item \textbf{Adaptive TL model for varying lighting conditions:} Unlike previous approaches that assume static lighting or require frequent manual recalibration, our model dynamically adjusts its learned representations to maintain high accuracy in fluctuating environments.

\item \textbf{Cost-efficient training and validation for faster model adaptation:} 
Moreover, our framework significantly reduces the data burden by requiring only 30\% of the data compared to traditional models. In contrast to existing methodologies that depend on large-scale data acquisition leading to higher costs and slower model updates, our approach enables faster and more cost-effective model adaptation, thereby facilitating rapid deployment in dynamic industrial environments.

\end{itemize}

The rest of the paper is organized as follows: Section~\ref{sec:sysmodel} presents the system model and problem formulation. Section~\ref{sec:algorithm} details the proposed algorithm development. In Section~\ref{sec:results}, we go through the performance evaluation and discussion before concluding in Section~\ref{sec:Conclusion}.

\section{System model and problem formulation}
\label{sec:sysmodel}

\begin{figure}[!t]
  \centering
  \resizebox{0.7\linewidth}{!}{\includegraphics{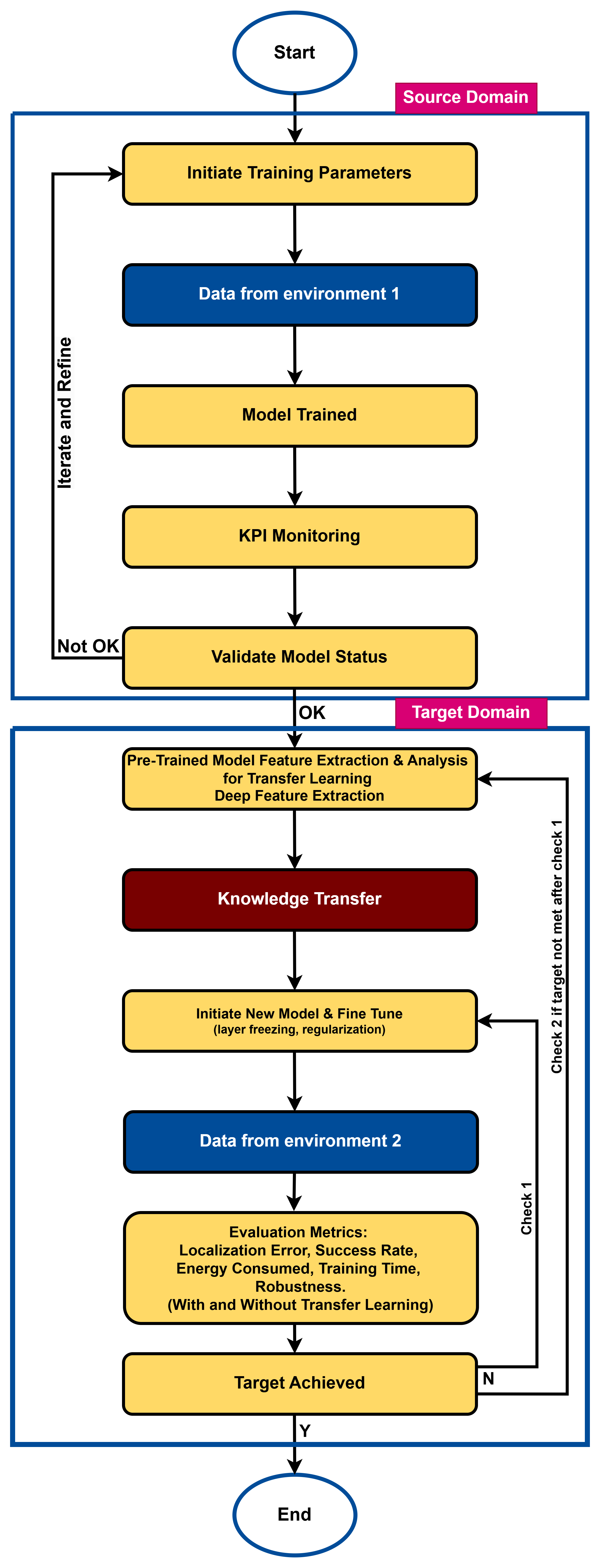}}
  \caption{Proposed Framework of TL-based localization process.}
  \label{fig:1}
\end{figure}

We aim to implement a TL-based approach using a DNN architecture to predict users' coordinates (longitude and latitude) based on collected VLC data. The approach, detailed in Fig.~\ref{fig:1}, uses the transfer of knowledge from a source model \( f_s \) to improve the learning of the target model \( f_t \), enabling better generalization to the new task. We conduct our testing across three scenarios representing different models configurations:

\begin{enumerate}
    \item \textbf{Base Model:} This is the original DNN model trained on the raw VLC dataset, serving as the benchmark for our experiments.
    \item \textbf{EV (Environmental Variation) Model:} This is the base model with added Noise Factor (NF) to simulate lighting variability, reflecting real-world
    uncertainty in signal strength.
    \item \textbf{TL Model:} For this model, we load the pre-trained base model to the EV model and fine-tune it. This process aims to improve degraded Key Performance Indicators (KPIs) by transferring learned features from the original scenario to better handle the noisy scenarios.
\end{enumerate}

\subsection{Source and Target Domains}

Let \( D_s = \{(x_i, y_i)\}_{i=1}^{n_s} \) be the source domain, where \( x_i \in \mathbb{R}^d \) are input feature vectors, and \( y_i \in \mathbb{R}^{m_s} \) are the corresponding output labels, representing longitude and latitude. Here, \( \mathbb{R}^d \) is the \( d \)-dimensional feature space (signal intensity values in our case), and \( \mathbb{R}^{m_s} \) is the space of the output labels, where \( m_s = 2 \) corresponds to the two output labels (longitude and latitude). The task \( T_s \) learns a function \( f_s: \mathbb{R}^d \rightarrow \mathbb{R}^{m_s} \), trained on the source domain.
Similarly, the target domain \( D_t = \{(x_j, y_j)\}_{j=1}^{n_t} \) has \( x_j \in \mathbb{R}^d \) and labels \( y_j \in \mathbb{R}^{m_t} \), with \( m_t = 2 \) (longitude and latitude), as in the source domain.
The goal is to learn \( f_t: \mathbb{R}^d \rightarrow \mathbb{R}^{m_t} \) that generalizes to the target task. The source model is a DNN with parameters \( \theta_s \), where:
\begin{equation}
    f_s(x_i; \theta_s) = \hat{y}_i,
\end{equation}
where \( \hat{y}_i \) is the predicted output 
in the source domain.

For the target task, the goal is to learn \( \theta_t \) for the target model, where the target function \( f_t \) is initialized from the pre-trained source model \( f_s \), and updated using the target data:
\begin{equation}
    f_t(x_j; \theta_t) = \hat{y}_j,
\end{equation}
\subsection{Loss Function}
We minimize a combined loss function that includes both the source domain loss \( \mathcal{L}_s \) and the target domain loss \( \mathcal{L}_t \):
\begin{equation}
    \mathcal{L}_{total} = \lambda_s \mathcal{L}_s(\theta_s) + \lambda_t \mathcal{L}_t(\theta_t),
\end{equation}
where \( \lambda_s \) and \( \lambda_t \) are weighting factors, optimized via cross-validation to enhance target domain performance. This ensures adaptation to the target domain while preserving useful source domain features. The source and target losses are computed as:
\begin{equation}
    \mathcal{L}_s = \frac{1}{n_s} \sum_{i=1}^{n_s} \mathcal{L}(f_s(x_i; \theta_s), y_i),
\end{equation}
\vspace{-0.5cm}
\begin{equation}
    \mathcal{L}_t = \frac{1}{n_t} \sum_{j=1}^{n_t} \mathcal{L}(f_t(x_j; \theta_t), y_j),
\end{equation}
where \( \mathcal{L} \) represents the loss function (Mean Squared Error (MSE)), and \( \mathcal{L}_s \) and \( \mathcal{L}_t \) are the individual losses computed for the source and target tasks, respectively.

\subsection{Fine-Tuning}

In the fine-tuning process, we initialize the target model with the pre-trained source parameters \( \theta_s \) and then update the target parameters \( \theta_t \) by applying small adjustments \( \Delta \theta \). These adjustments specialize the model for the target task while retaining valuable features learned from the source domain, thereby improving performance by reducing the target domain's loss \( \mathcal{L}_t \) and enhancing generalization. Specifically, the target parameters are updated as:

\begin{equation}
    \theta_t = \theta_s + \Delta \theta,
\end{equation}
\section{Algorithm development and implementation}
\label{sec:algorithm}

\subsection{Pre-TL Model Development}
Before applying TL for real-time localization, we first train the base model on the original dataset collected at time \( t \) and ensure that it reaches a satisfying performance. 
\begin{equation}
\mathcal{L}_{train} = \frac{1}{|\mathcal{D}_{train}|} \sum_{i=1}^{|\mathcal{D}_{train}|} | \hat{y}_i - y_i |, 
\end{equation}
\begin{equation}
\mathcal{L}_{val} = \frac{1}{|\mathcal{D}_{val}|} \sum_{i=1}^{|\mathcal{D}_{val}|} | \hat{y}_i - y_i |,
\end{equation}
where \(\mathcal{L}_{train}\) and \(\mathcal{L}_{val}\) represent the training and validation localization errors, respectively, and \(\hat{y}_i\) and \(y_i\) are the predicted and true localization values (longitude and latitude) for each data point. The datasets \(\mathcal{D}_{train}\) and \(\mathcal{D}_{val}\) represent the training and validation sets, respectively, containing pairs of input features (signal intensities) and corresponding localization values. After training, the model parameters \(W_r\) are updated. 

As previously mentioned, localization models in indoor environments require periodic updates to adapt to environmental changes over time. Therefore, after training the base model and obtaining the necessary metrics for benchmarking, we simulate environmental variability into the base model by introducing Gaussian noise to the original dataset. The spread of the Gaussian noise is controlled by the standard deviation \(\sigma\), creating a noisy version of the original dataset that reflects conditions at time \( t + i \), where \( i \) represents the collection time. We refer to the models trained in these environments as EV models, where each EV model corresponds to a specific noisy environment with a NF of 2, 4, or 8, capturing different environmental scenarios as follows:
\begin{equation}
\mathcal{D}_{train}^{noisy} = \mathcal{D}_{train} + \mathcal{N}(0, \sigma^2),
\label{eq:1}
\end{equation}
where \(\mathcal{N}(0, \sigma^2)\) denotes Gaussian noise with zero mean and variance \(\sigma^2\). The NF scales the standard deviation of the noise as follows:
\begin{equation}
\sigma = NF \cdot \sigma_{base},
\label{eq:2}
\end{equation}
where \(\sigma_{base}\) is the baseline standard deviation. 
The model is updated using Stochastic Gradient Descent (SGD), which iteratively minimizes the loss function to improve model parameters:
\begin{equation}
W_{r+1} = W_r - \eta \nabla L(W_r),
\label{eq:3}
\end{equation}
where \(L(W_r)\) represents 
loss function and \(\eta\) is 
learning rate. 

As a result, the EV model experiences degradation in KPIs compared to the base model. This performance decline, underscores the necessity of employing TL to effectively adapt the model to the evolving environmental conditions while preserving its localization accuracy.

\subsection{TL Algorithm Development and Implementation}

Similar to the base and EV models, our TL model is constructed using the TensorFlow Keras API, ensuring flexibility and efficiency in the environment and fine-tuning DNN model \cite{chicho2021comprehensive}. The approach aims to transfer learned knowledge from the base model to the EV model, enabling accurate localization predictions while optimizing computational resources.


\begin{algorithm}
\scriptsize 
\caption{Transfer Learning for Localization}
\SetAlgoLined
\SetKwInOut{Input}{Input}
\SetKwInOut{Output}{Output}

\Input{$\mathcal{D}_{train}$; /* User VLC dataset */}
\Input{$\mathcal{D}_{val}$; /* User VLC Dataset */}
\Input{$\alpha$; /* Output weights */}
\Input{$\sigma^2$; /* Variance of Gaussian noise */}
\Output{$W$; /* Trained model parameters */}

\BlankLine

\SetKwProg{ServerInit}{ServerInit}{}{}
\ServerInit{}{
    Initialize $W_0$, $r \gets 0$ \;
}

\SetKwProg{LoadBaseModel}{Load Base Model}{}{}
\LoadBaseModel{}{
    Load $\mathcal{M}_{base}$ and its weights $W_{base}$;
}

\SetKwProg{DefineTLModel}{Define TL Model}{}{}
\DefineTLModel{}{
    Define $\mathcal{M}_{TL}$;
}

\SetKwProg{TransferWeights}{Transfer Weights}{}{}
\TransferWeights{}{
    Transfer $W_{base}$ to $\mathcal{M}_{TL}$ and set shared layer weights;
}

\SetKwProg{FreezeSharedLayers}{Freeze Shared Layers}{}{}
\FreezeSharedLayers{}{
    \ForEach{layer in $\mathcal{M}_{TL}$}{
        \If{layer is in shared layers}{
            Freeze the layer by setting `layer.trainable = False`;
        }
    }
}

\SetKwProg{FineTuneModel}{Fine-Tuning}{}{}
\FineTuneModel{}{
    Fine-tune the model with $\mathcal{D}_{train}$, updating only the non-frozen layers;
    Update $W_{r}$ using backpropagation;
}

\While{$r < r_{max}$ and not converged}{
    \SetKwProg{TrackEpochTime}{Track Epoch Time}{}{}
    \SetKwProg{AddNoise}{Simulate EV Model 
    }{}{}
    \AddNoise{}{
        $\mathcal{D}_{train}^{noisy} = \mathcal{D}_{train} + \mathcal{N}(0, \sigma^2)$;
    }

    \TrackEpochTime{}{
        $T_{start} \gets \text{start time}$\;
    }

    \SetKwProg{TrainModel}{Train Model}{}{}
    \TrainModel{}{
        Update model with $W_r$\;
        Train model using:
        $\nabla L(W_r) = \frac{1}{|\mathcal{D}_{train}^{noisy}|} \sum_{i=1}^{|\mathcal{D}_{train}^{noisy}|} \nabla f(x_i, y_i, W_r)$\;
        Update weights using:
        $W_{r+1} = W_r - \eta \nabla L(W_r)$ \;
    }

    \SetKwProg{TrackEpochTimeEnd}{Track Epoch Time End}{}{}
    \TrackEpochTimeEnd{}{
        $T_{end} \gets \text{end time}$\;
        $T_{epoch} = T_{end} - T_{start}$\;
    }

    \SetKwProg{EnergyConsumed}{Monitor Energy Consumed}{}{}
    \EnergyConsumed{}{
        $E_{epoch} = (P_{GPU} + P_{CPU}) \cdot T_{epoch}$\;
        $E_{total} += E_{epoch}$; /* Update cumulative energy */;
    }

    \SetKwProg{LocalizationError}{Compute Localization Error}{}{}
    \LocalizationError{}{
        $\mathcal{L}_{train} = \frac{1}{|\mathcal{D}_{train}|} \sum | \hat{y}_i - y_i |$;
        $\mathcal{L}_{val} = \frac{1}{|\mathcal{D}_{val}|} \sum | \hat{y}_i - y_i |$;
    }

    \SetKwProg{SuccessRate}{Compute Success Rate}{}{}
    \SuccessRate{}{
        $SR_{train} = \frac{|\mathcal{D}_{train}|_{\mathcal{L}_{train} < \delta}}{|\mathcal{D}_{train}|}$, 
        $SR_{val} = \frac{|\mathcal{D}_{val}|_{\mathcal{L}_{val} < \delta}}{|\mathcal{D}_{val}|}$;
    }

    \SetKwProg{SaveMetrics}{Save Metrics}{}{}
    \SaveMetrics{}{
        Save metrics: $\mathcal{L}_{train}$, $\mathcal{L}_{val}$, $SR_{train}$, $SR_{val}$, $E_{con}$, $T_{epoch}$;
    }
}

Save model: $transfer\_learning\_model.save.h5$\;
\end{algorithm}

We design the TL algorithm to take as input a set of received signal strength dataset denoted as $D_{train}$ and $D_{val}$, a set of output weights $\alpha$, and the variance of Gaussian noise $\sigma^2$. The trained model parameters, represented as $W$, will be the output. The algorithm starts by initializing the server, setting the initial model parameters $W_0$ and a variable $r$ to track the number of epochs. The base model \( \mathcal{M}_{base} \), along with its weights \( W_{base} \), is loaded, and a TL model \( \mathcal{M}_{TL} \) is defined. The weights \( W_{base} \) are transferred to \( \mathcal{M}_{TL} \), and the shared layer weights (layers reused from the base model) are set accordingly. 
Fine-tuning is performed using the training dataset \( \mathcal{D}_{train} \), updating only the non-frozen layers via backpropagation to obtain the new weights \( W_{r+1} \).

Following the methodology described in Section~III.A, a similar EV model scenario is simulated by introducing Gaussian noise into the original training dataset. 
This EV model is created to facilitate the application of TL, allowing the knowledge from the base model to be transferred and fine-tuned to the EV model in this TL framework. During each communication epoch \( r \), provided that the maximum epoch \( r_{max} \) is not reached and the convergence criterion is unmet, the algorithm performs several steps. The epoch time \( T_{epoch} \) is tracked as the difference between the start and end times of each epoch. 
The model is updated with the current weights \( W_r \), and training is performed using gradient descent. The loss gradient \( \nabla L(W_r) \) is calculated as the average gradient over the noisy training dataset, and the weights are updated as \( W_{r+1} = W_r - \eta \nabla L(W_r) \), where \( \eta \) is the learning rate.

Energy consumption during training is monitored by computing the energy used in each epoch as \( E_{epoch} = (P_{GPU} + P_{CPU}) \cdot T_{epoch} \), where \( P_{GPU} \) and \( P_{CPU} \) represent power consumption of the GPU and CPU, respectively. The cumulative energy \( E_{total} \) is updated after each epoch. Localization error for training and validation datasets is computed as \( \mathcal{L}_{train} \) and \( \mathcal{L}_{val} \), defined as the average absolute difference between predicted and true localization values. Success rates \( SR_{train} \) and \( SR_{val} \) are calculated as the fraction of data points with localization error less than or equal to \( \delta \), where \( \delta = 1 \) meter. These metrics, along with the energy consumed \( E_{con} \), are saved after each communication epoch. The process repeats until convergence or the maximum number of communication epochs is reached.

\section{Performance Evaluation and Discussion} 
\label{sec:results}

\subsection{VLC Measurement System}

The VLC measurement system consists of two LED transceivers, two converters, two Direct Current (DC) power sources, and a vector network analyzer (VNA). In this setup, the LED transmitter and the Photo-Diode (PD)-based receiver are integrated into a single hardware unit. The PD features a semi-angle at half sensitivity of 62 degrees and a 3-dB bandwidth of 50 MHz. The converter combines the swept frequency signal with the DC power supply or isolates them at the network port, enabling both information transmission and energy delivery through a single network cable. Furthermore, the VNA is configured with port 1 and port 2 connected to the transmitter and receiver, respectively, ensuring accurate signal analysis.


\subsection{Data Collection} 

\begin{figure}[h]
  \centering
  \begin{subfigure}[b]{0.6\linewidth} 
    \centering
    \includegraphics[width=\linewidth]{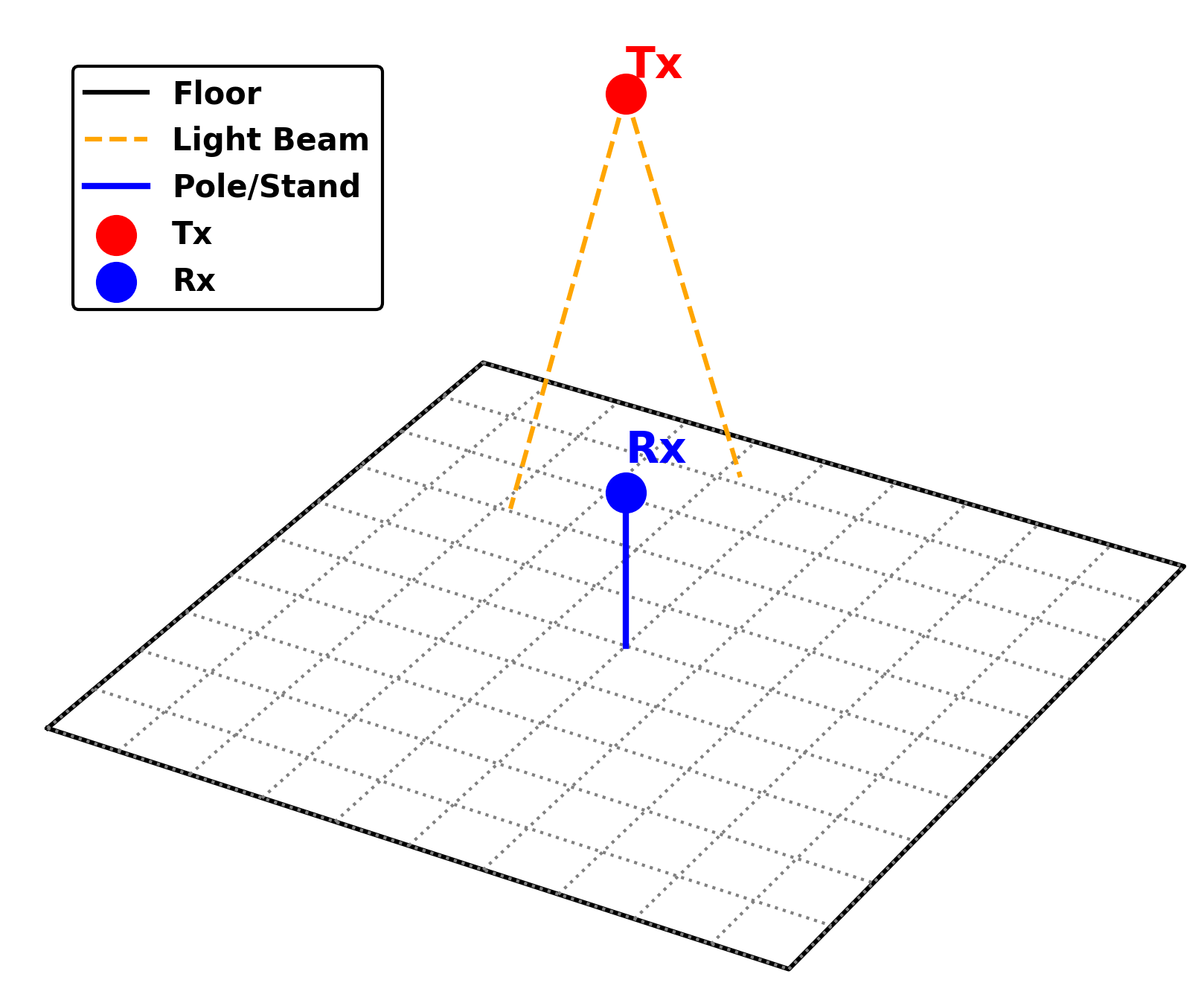}
    \caption{} 
    \label{fig:2a}
  \end{subfigure}
  \vspace{0.5cm} 
  \begin{subfigure}[!t]{0.9\linewidth} 
    \centering
    \includegraphics[width=\linewidth]{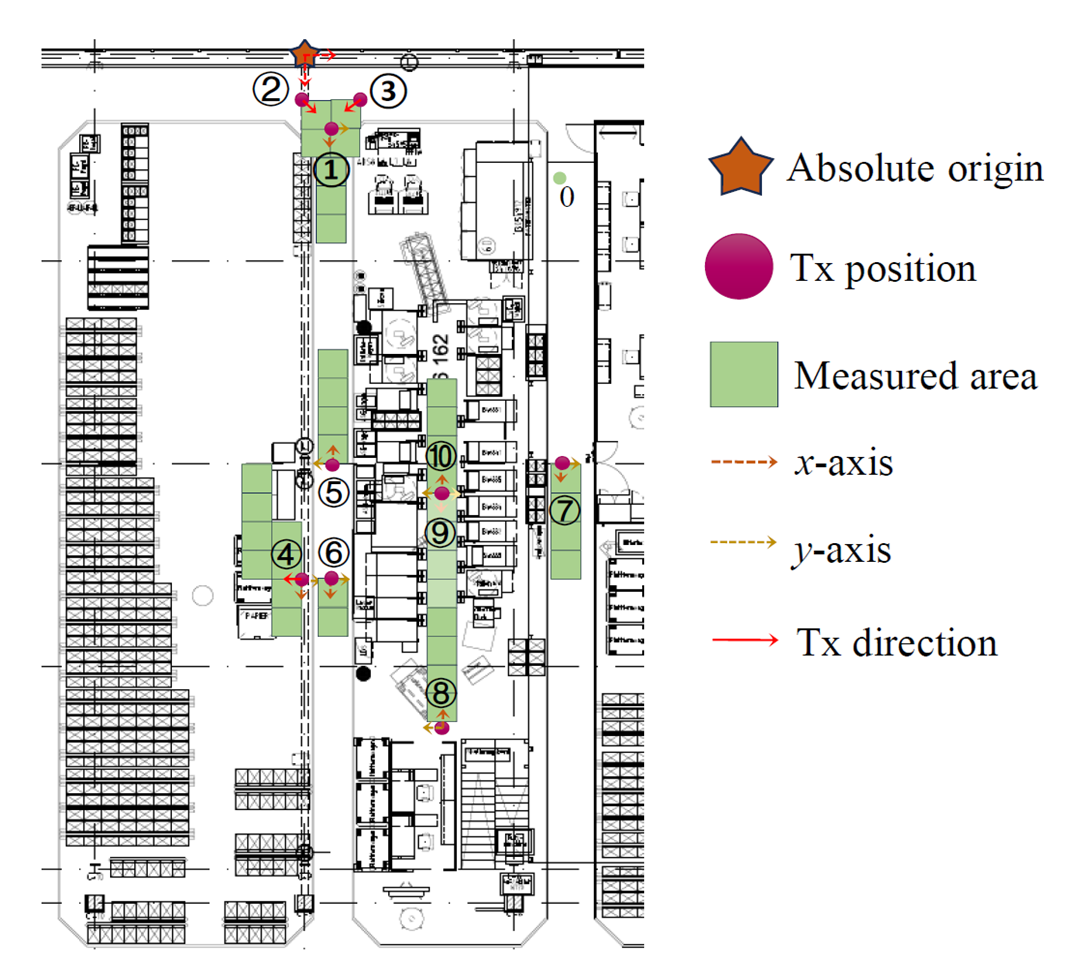}
    \caption{} 
    \label{fig:2b}
  \end{subfigure}
  \caption{(a) The concept of the transceiver deployment and grid distribution, and (b) the location of the transmitter and measured area in the production line.}
  \label{fig:2}
\end{figure}

The VLC data collection is performed at one of the production lines of the BOSCH factory in Blaichach, Germany. Fig.~\ref{fig:2}(a) illustrates the sketch of the transceiver deployment and grid distribution. Based on the practical conditions of the factory environment and the operating platform of the production line, the heights of the transmitter front-end and receiver front-end are fixed at 2.9 m and 1.1 m, respectively. 
For systematic analysis and ease of statistical processing, the receiving plane was divided into evenly spaced grids, each measuring 0.2 × 0.2 m², and the intersection of each grid was used as a measurement point for data collection.
Fig.~\ref{fig:2}(b) depicts the practical location of the transmitter and measured area in the production line scenario.  The orange star defines the absolute origin during the VLC data measurements. We put 10 LEDs based on the practical conditions of the production line and therefore 9 measurement scenarios are considered in the VLC data collection. The average number of receiver positions across each scenario is 160. At each receiver position, measurements were recorded for 5 seconds, and the average of the measured values was taken to obtain a single Received Signal Strength Indicator (RSSI) value for each receiver position.

Specifically, scenarios 1 and 2 (point 1,2,3) are close to both the center console of the production line and the transparent windows. Scenario 3 (point 4) is located in a material storage area. 
Scenarios 4 and 5, which correspond to point 5 and 6 respectively, are both close to the back of the production line machine, while the left side of the scenarios 4 is bordered two cement pillars and the left side of the scenarios 5 is adjacent to the material storage scenario. Moreover, there is a row of lockers and 
conveyor belts on the left and at the top of the scenario 6 (point 7 ). Note that the height of the transmitter is higher than the conveyor belts. Furthermore, scenarios 7-9 (point 8-10) are located in the middle of the production line, where workers mainly move around and control the operating console. 
The precise coordinates of the transmitter in these scenarios is listed in TABLE~\ref{Table:1}.

\begin{table}[!t]
\centering
\caption{3D Coordinates of the TX Positions.}
\renewcommand{\arraystretch}{1.2} 
\scriptsize 
\begin{tabular}{|c|c|c|c|}
\hline
\textbf{Transmitter ID} & \textbf{Position (x, y, z)} & \textbf{Transmitter ID} & \textbf{Position (x, y, z)} \\
\hline
TX1 & (2.6, 0.8, 2.9) & TX6 & (17.1, 0.8, 2.9) \\
\hline
TX2 & (1.7, -0.2, 2.9) & TX7 & (14.2, 8.5, 2.9) \\\hline
TX3 & (3.7, 1.8, 2.9) & TX8 & (23.2, 4.6, 2.9) \\\hline
TX4 & (17.1, 0.2, 2.9) & TX9 & (15.2, 4.6, 2.9) \\\hline
TX5 & (12.9, 0.8, 2.9) & TX10 & (13.5, 3.6, 2.9) \\
\hline
\end{tabular}
\label{Table:1}
\end{table}

\subsection{TL Model Evaluation under different lighting conditions}

We evaluate the performance of three models: the Base model, the EV model, and the TL model. All three models use a DNN with five hidden layers. We employ the ADAM optimizer with a learning rate of 0.0001 and use the MSE as the loss function. We adopt an 80/20 training-validation split for all experiments. Training is conducted for 600 epochs with a batch size of 64. TABLE~\ref{Table:5} summarizes all experimental parameters.

We firstly train the base model on the VLC dataset and optimize the algorithm to get the best possible results via an exhaustive process of simulations. 
Then, we explore a range of NFs to simulate an environment where the lighting characteristics are changing. For our experiments, we select NFs $\in \{2, 4, 8\}$. Finally we evaluate our TL model by loading the pre-trained DNN models.

\begin{table}[!t]
\caption{TL Simulation Settings.}
    \centering
    \scalebox{1}{
    \begin{tabular}{ccc}
        \toprule
        \textbf{Parameter} & \textbf{Description} & \textbf{Value} \\
        \midrule
        \textit{d} & Input dimension & 10 \\
        Optimizer & Model optimizer & Adam \\
        $\eta$ & Learning rate & 0.0001 \\
        \textit{B} & Batch size & 64 \\
        \textit{E} & Number of epochs per iteration & 600 \\
        \textit{L} & Number of hidden layers & 5 \\
        \textit{h} & Max size of hidden layer & 512 \\
        \bottomrule
    \end{tabular}
    }
    \label{Table:5}
\end{table}

\begin{figure*}[!t]
    \centering
    \begin{minipage}{0.32\textwidth}
        \centering
        \includegraphics[width=\linewidth]{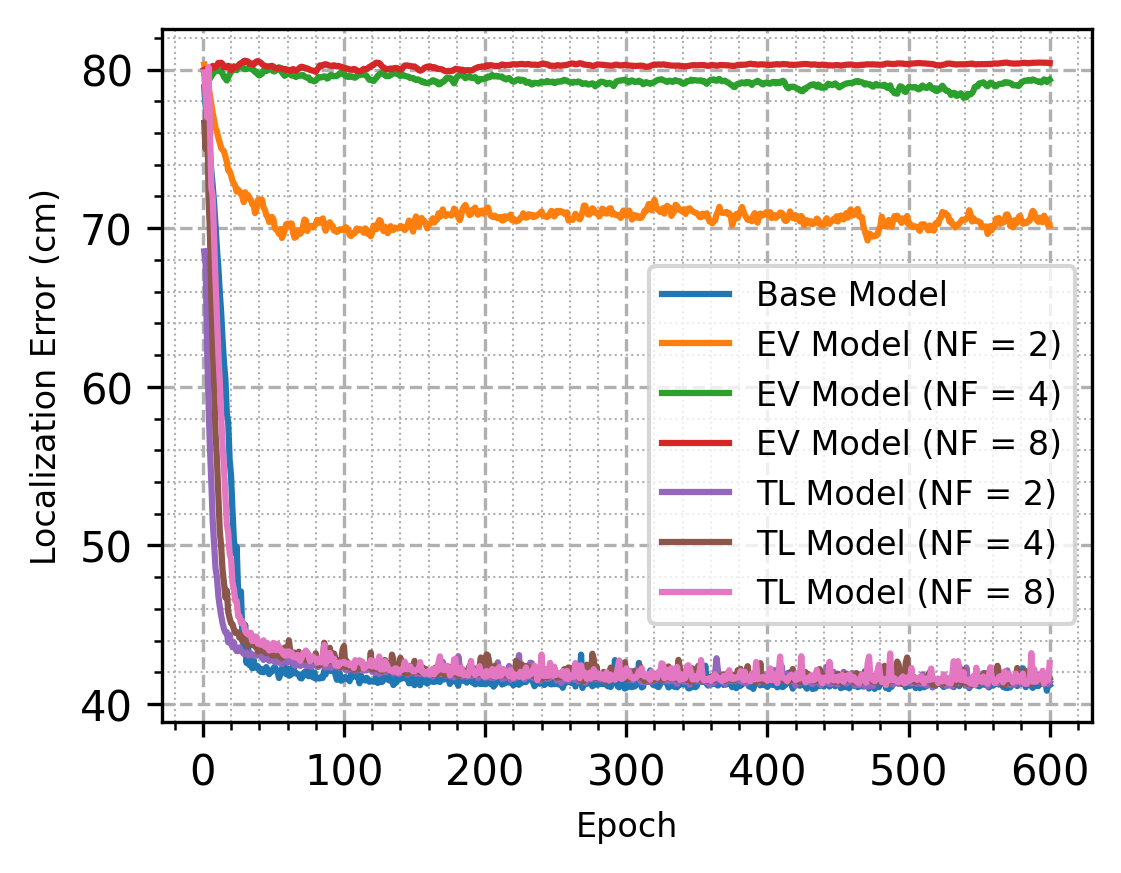}
        \subcaption{Localization Error.}
        \label{fig:loc_error}
    \end{minipage}
    \hfill
    \begin{minipage}{0.32\textwidth}
        \centering
        \includegraphics[width=\linewidth]{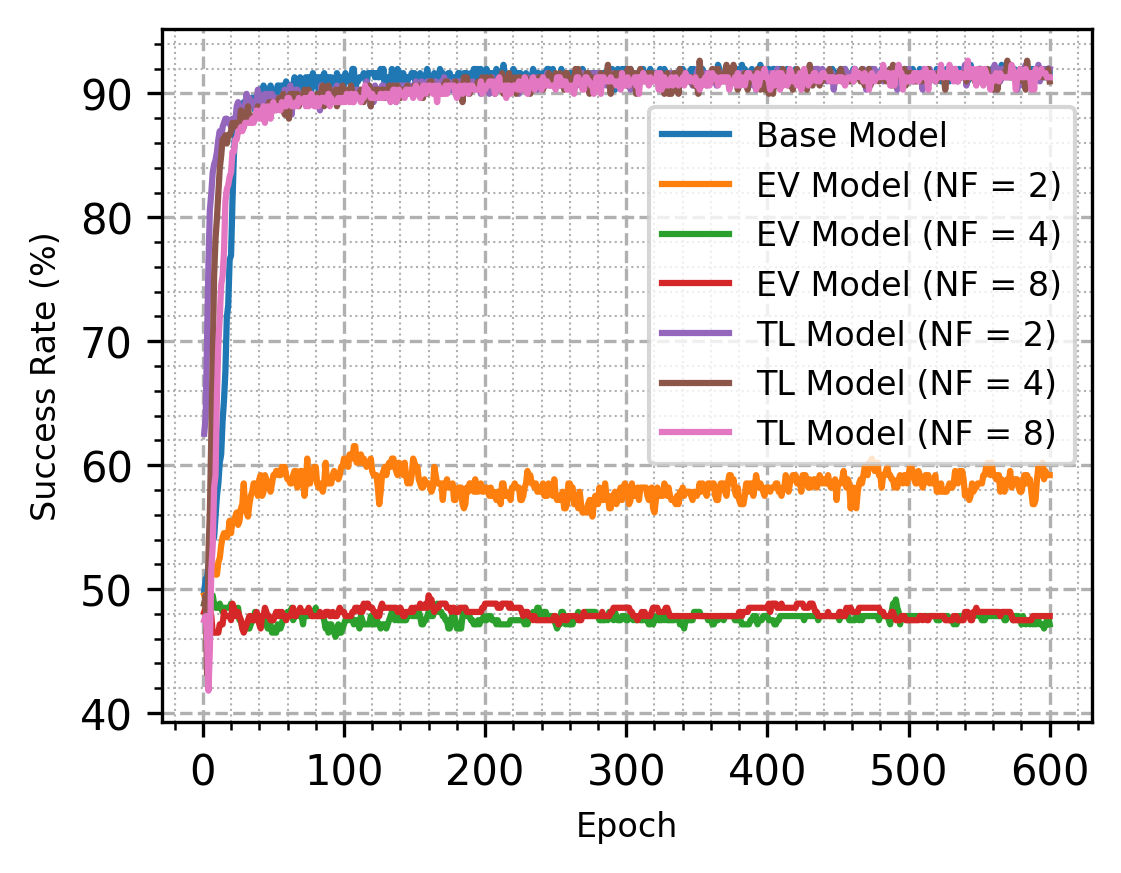}
        \subcaption{Success Rate.}
        \label{fig:success_rate}
    \end{minipage}
    \hfill
    \begin{minipage}{0.32\textwidth}
        \centering
        \includegraphics[width=\linewidth]{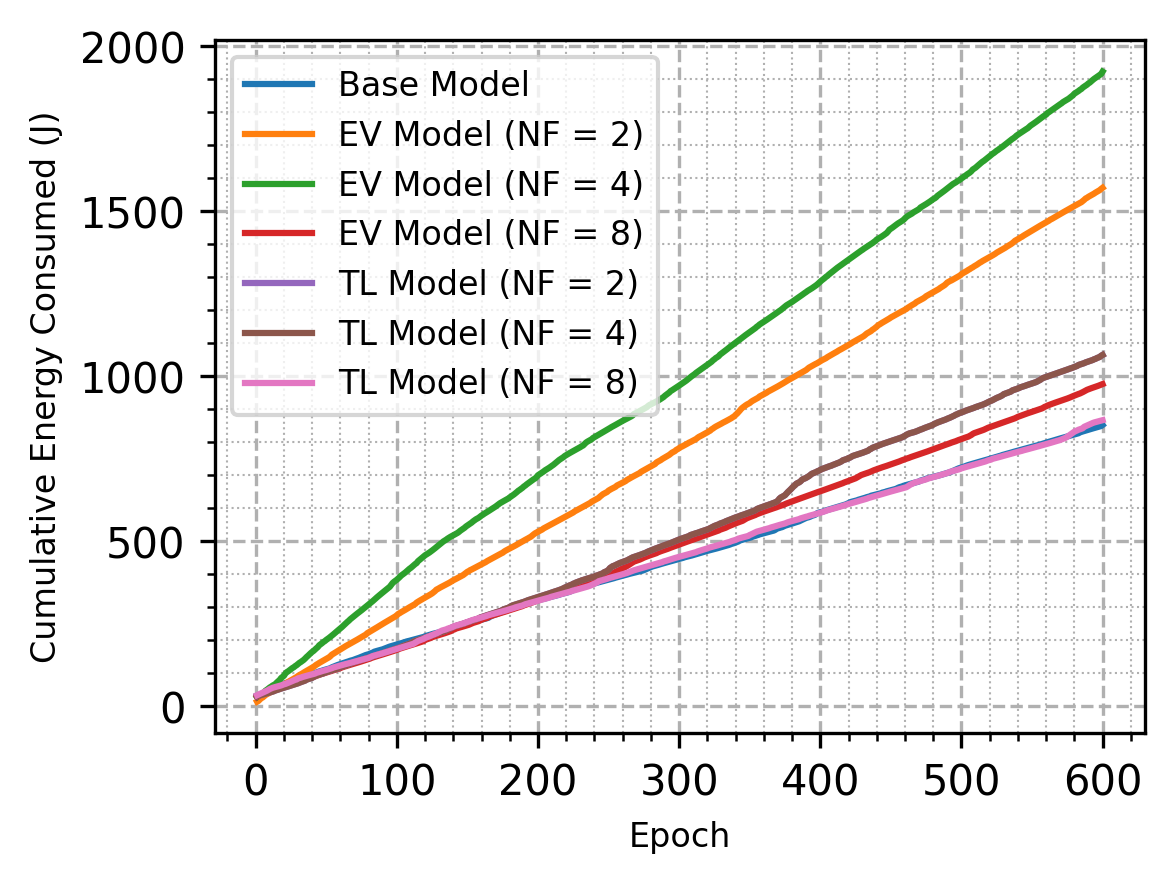}
        \subcaption{Energy Consumed.}
        \label{fig:energy_consumed}
    \end{minipage}
    
    \vspace{0.5cm} 
    
    \begin{minipage}{0.32\textwidth}
        \centering
        \includegraphics[width=\linewidth]{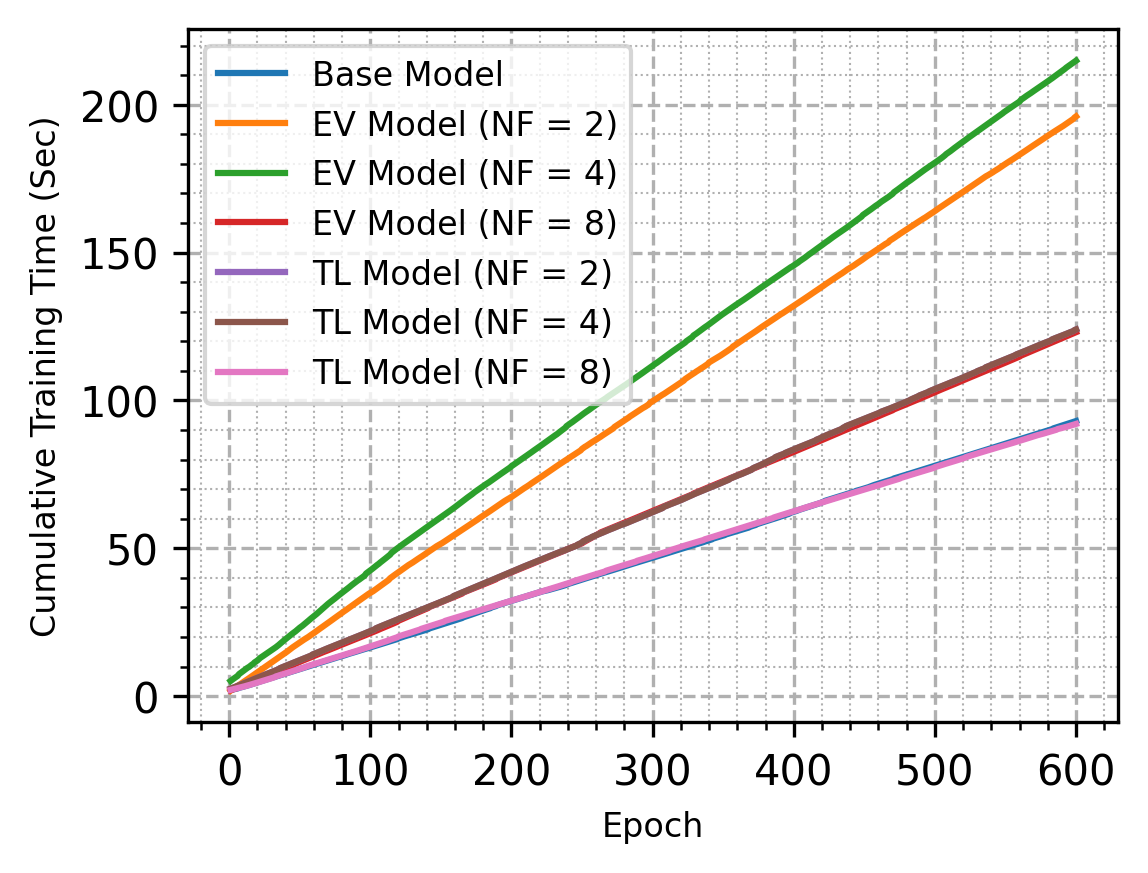}
        \subcaption{Training Times.}
        \label{fig:training_time}
    \end{minipage}
    \hfill
    \begin{minipage}{0.32\textwidth}
        \centering
        \includegraphics[width=\linewidth]{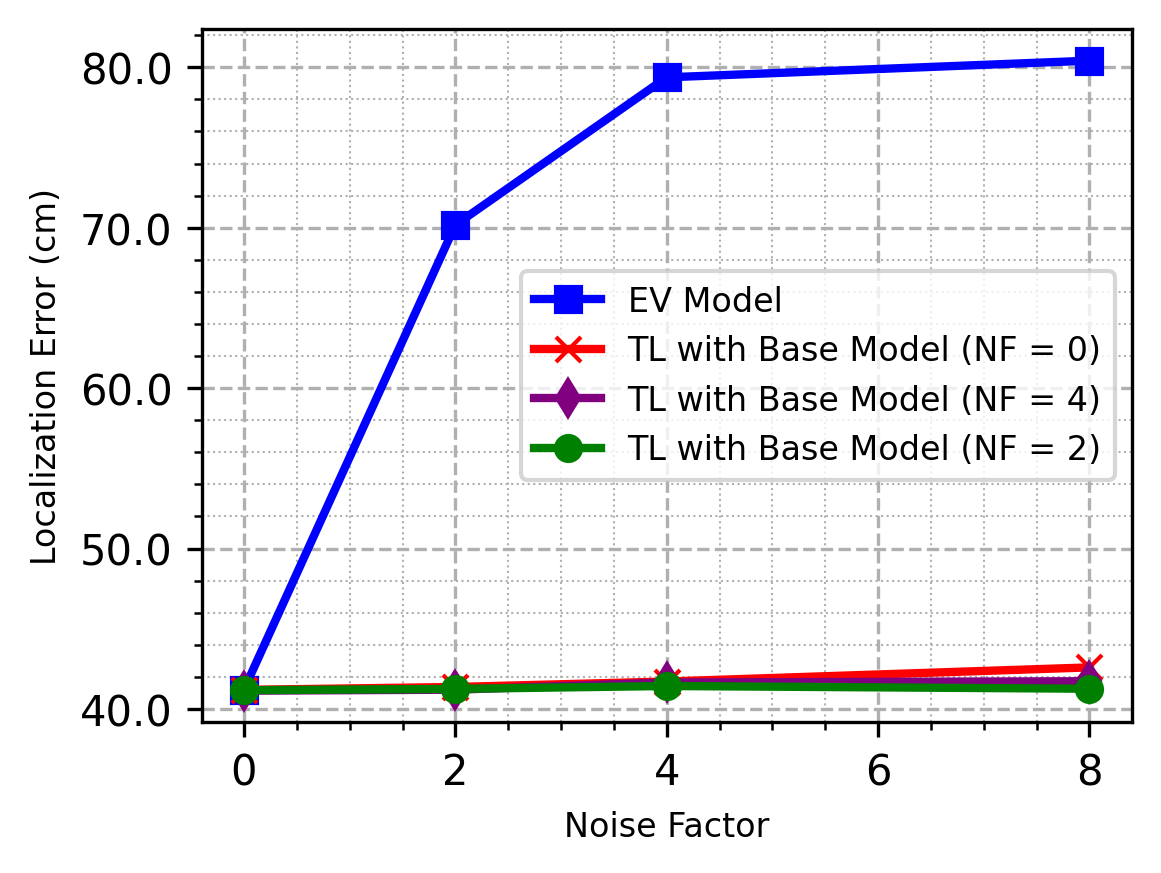}
        \subcaption{TL of Base Model with different NFs.}
        \label{fig:learning_curve}
    \end{minipage}
    \hfill
    \begin{minipage}{0.32\textwidth}
        \centering
        \includegraphics[width=\linewidth]{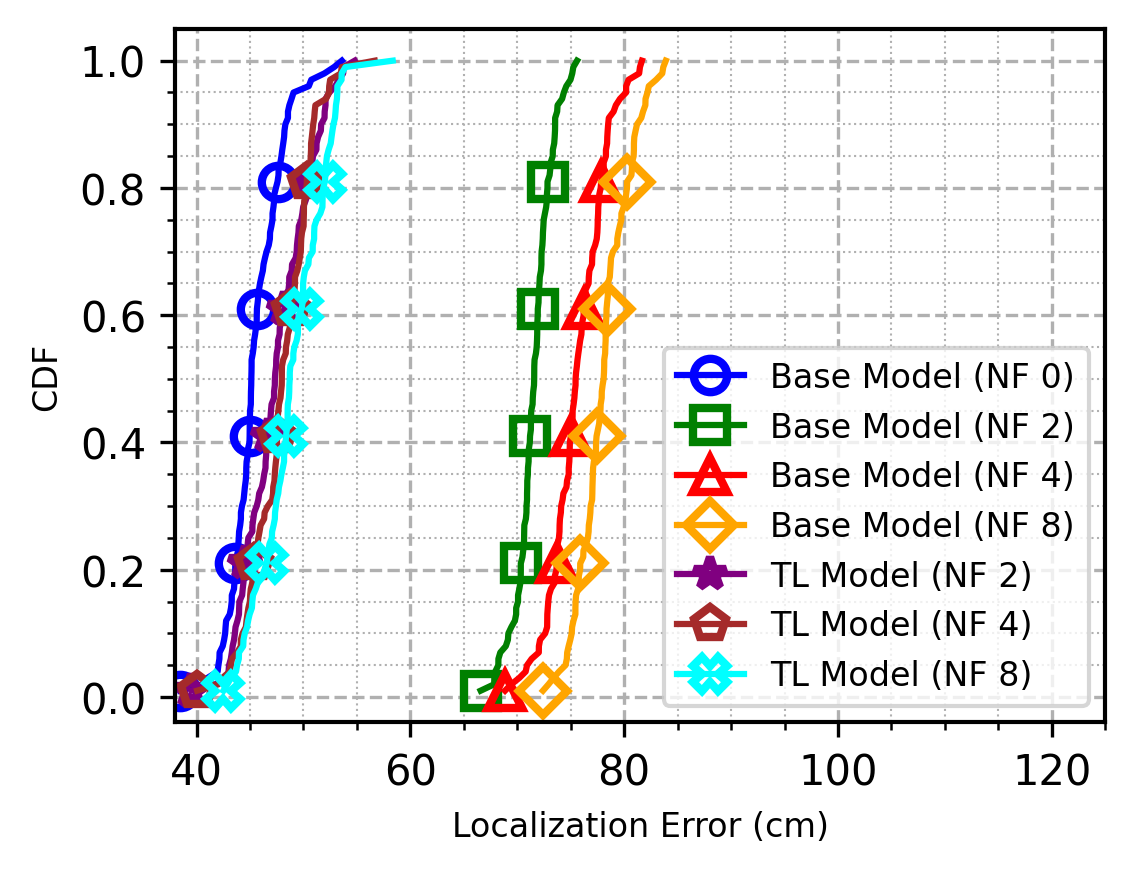}
        \subcaption{CDF Plots.}
        \label{fig:cdf_plots}
    \end{minipage}
    
    \caption{TL performance evaluation.}
    \label{fig:performance_eval}
\end{figure*}

\begin{table*}[!t] 
\centering
\caption{Comparison of final metrics across models with different NFs, using the base model at NF = 0.}
\label{tab:final_metrics_combined}
\begin{tabular}{|l|c|c|c|c|c|c|c|c|}
\hline
\textbf{Final Metric Value} & \textbf{NF = 0} & \multicolumn{2}{c|}{\textbf{NF = 2}} & \multicolumn{2}{c|}{\textbf{NF = 4}} & \multicolumn{2}{c|}{\textbf{NF = 8}} \\
\hline
\textbf{{Training Model}} & \textbf{Base Model} & \textbf{EV Model} & \textbf{TL Model} & \textbf{EV Model} & \textbf{TL Model} & \textbf{EV Model} & \textbf{TL Model} \\
\hline
\text{Training Localization Error (cm)} & 41.13 & 70.38 & 41.37 & 78.99 & 41.74 & 79.91 & 42.42 \\
\hline
\text{Validation Localization Error (cm)} & 41.21 & 70.17 & 41.37 & 79.37 & 41.71 & 80.41 & 42.59 \\
\hline
\text{Training Success Rate (\%)} & 93.07 & 57.76 & 93.07 & 47.25 & 92.32 & 47.75 & 92.49 \\
\hline
\text{Validation Success Rate (\%)} & 91.64 & 59.20 & 91.97 & 47.16 & 90.97 & 47.83 & 91.30 \\
\hline
\text{Cumulative Energy Consumed (J)} & 852.79 & 1571.35 & 1064.91 & 1923.38 & 1082.56 & 976.70 & 867.22 \\
\hline
\text{Cumulative Training Time (s)} & 93.04 & 196.03 & 124.00 & 214.88 & 126.17 & 123.08 & 92.06 \\
\hline
\text{Training Time per Epoch (s)} & 0.21 & 0.35 & 0.19 & 0.32 & 0.17 & 0.20 & 0.15 \\
\hline
\end{tabular}
\label{Table:2}
\end{table*}

\begin{table*}[!t]
\caption{Localization error comparison of TL models at different NFs.}
\centering
\renewcommand{\arraystretch}{1.0}
\begin{tabular}{|c|c|c|c|c|}
\hline
\textbf{NF} & \textbf{EV Model (cm)} & \textbf{TL as Base Model (cm)} & \textbf{TL as (EV Model NF =2) (cm)} & \textbf{TL as (EV Model NF =4) (cm)} \\
\hline
0 & - & 41.19 & 41.17 & 41.14 \\
\hline
2 & 70.17 & 41.37 & 41.25 & 41.20 \\
\hline
4 & 79.37 & 41.71 & 41.43 & 41.65 \\
\hline
8 & 80.41 & 42.59 & 41.26 & 41.73 \\
\hline
\end{tabular}
\captionsetup{justification=centering} 
\label{tab:localization_error}
\label{Table:3}
\end{table*}

Fig. 3(a) represents the localization error during the validation step. As indicated in Table~\ref{Table:2}, the base model achieves the best performance, with a localization error of 41.21 cm. 
However, the EV model demonstrates significant sensitivity to noise, with localization errors increasing from 40.21 cm to 70.17 cm at $NF = 2$, and further rising to 80.41 cm at $NF = 8$. This represents a 95.14\% increase compared to the base model's error. In contrast, the TL model effectively mitigates the impact of noise, maintaining low localization errors of 41.37 cm and 42.59 cm at $NF = 2$ and $NF = 8$, respectively. These correspond to minimal increases of only 0.39\% and 3.35\% over the base model. While our localization error is higher than some existing methods, our aim is to represent a calculated trade-off for a robust, energy-efficient TL framework that mitigate noise and performs well with limited data. Applying TL to more sophisticated models than DNN could further reduce localization error, but with increased computational complexity and energy demands.

Fig. 3(b) highlights the success rate, where the EV model demonstrates a steep decline, dropping below 50\% at higher noise levels, i.e., $NF \in \{4,8\}$. This sensitivity contrasts with the TL model, which preserves a high success rate of over 91\%, with minor reductions of up to 0.73\%, compared to the base model. Fig. 3(c) focuses on cumulative energy consumption. 
The TL model demonstrates a notable improvement, 
reflecting a decrease of 54.91\% compared to the EV model under the same noise conditions. Fig. 3(d) illustrates cumulative training time, where 
the TL model drastically improves efficiency, reducing training time to 123.08 s, representing a reduction of 42.71\%, compared to the EV model.

We further evaluated the performance of our TL model by comparing it with TL models trained using different base models, as shown in Fig. 3(e), to assess how the choice of the base model impacts localization accuracy. TABLE~\ref{Table:3} shows that the TL model trained with base model ($NF = 2$) achieves a marginal improvement of 0.29\% localization error compared to the TL model trained with base model ($NF = 0$). Similarly, for $NF = 4$, the localization error improves by 0.67\% from the baseline TL model trained. These results indicate that while using a base model already trained with a NF does result in a slight reduction in localization error, however, the improvement is minimal, suggesting limited sensitivity to the choice of the base model trained on different NFs. 

To comprehensively evaluate localization accuracy, we generated the cumulative distribution functions (CDFs) for the localization error performance of various models, including the base model and the TL model, across different $NFs$ $(0, 2, 4, 8)$. These CDFs illustrate the distribution of localization errors, providing insights into the models’ ability to predict locations across the dataset accurately Fig. 3(f). 
The overall trend highlights the degradation in localization accuracy as noise levels increase, however, the TL models helps mitigate these noise effects.

\subsection{TL model evaluation with limited data}

In this part, we evaluate the TL model performance by comparing the localization errors under varying data sizes and noise levels. As shown in TABLE~\ref{Table:4}, reducing the data size to 30\% of the original size results in a significant increase in localization error, reaching 86.78 cm. However, after applying TL, the error decreases to approximately 42.96 cm, representing a reduction of over 45\%. This substantial improvement highlights the TL model's ability to perform effectively even with limited data. Fig. 4 illustrates that, after applying the TL approach, the localization error with 30\% of the data size closely matches to that of the full size dataset.

\begin{table}[!t]
\caption{Error values before and after TL at different NFs, comparing performance across various data sizes.}
\centering
\resizebox{\columnwidth}{!}{%
\begin{tabular}{|c|c|c|c|c|c|c|}
\hline
\textbf{Data Size} & \multicolumn{3}{c|}{\textbf{Error Before TL (cm)}} & \multicolumn{3}{c|}{\textbf{Error After TL (cm)}} \\
\hline
& \textbf{NF = 2} & \textbf{NF = 4} & \textbf{NF = 8} & \textbf{NF = 2} & \textbf{NF = 4} & \textbf{NF = 8} \\
\hline
30\% & 77.08 & 83.08 & 86.78 & 41.89 & 42.05 & 42.96 \\
\hline
50\% & 76.49 & 81.89 & 84.75 & 41.71 & 41.65 & 41.89 \\
\hline
70\% & 74.66 & 80.17 & 81.33 & 41.63 & 41.45 & 41.59 \\
\hline
100\% & 70.17 & 79.01 & 80.76 & 41.45 & 41.32 & 41.33 \\
\hline
\end{tabular}%
}
\label{Table:4}
\end{table}

\begin{figure}[!t]
    \centering
    \includegraphics[scale=0.85]{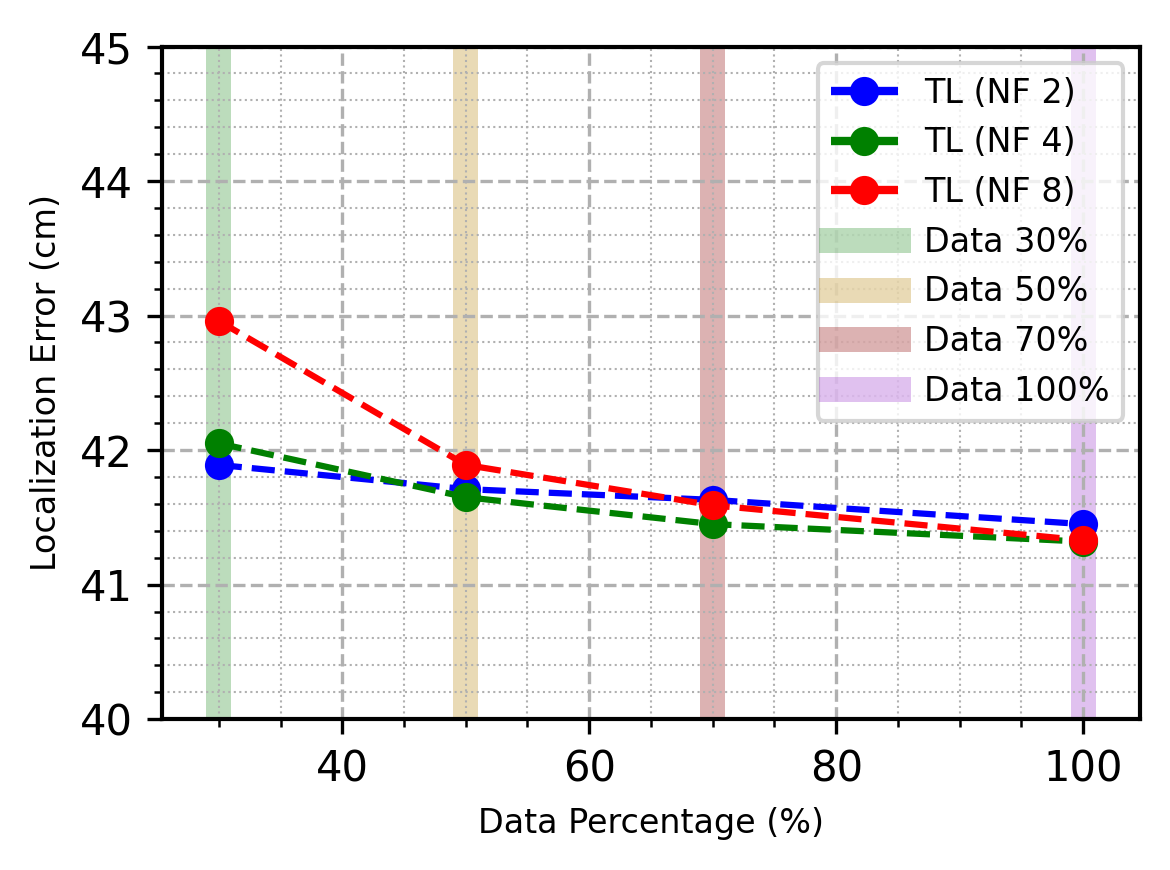}
    \caption{TL model evaluation with limited data.}
    \label{fig:g}
\end{figure}

\section{Conclusion}
\label{sec:Conclusion}

This paper addresses the challenge of reliable indoor localization in industrial dynamic environments. We propose a TL approach using VLC-based localization, combining deep learning with knowledge transfer to enhance localization performance under environmental changes. The TL model significantly improves localization accuracy, reduced errors by up to 47\% compared to traditional DNN methods. Furthermore, it achieves energy efficiency and reduces computational time by over 40\%, even in scenarios with fluctuating lighting conditions. 
Notably, the TL framework effectively adapts to new environments with minimal data requirements, using only 30\% of the dataset without compromising accuracy. 
Future research will focus on broadening the applicability of this framework by accounting for additional environmental variables, integrating diverse sensor modalities, and tackling issues such as computational scalability and real-time implementation in intricate industrial settings. 

\section{Acknowledgment}

The authors gratefully acknowledge the financial supports of the EU Horizon 2020 program towards the 6G BRAINS project H2020-ICT 101017226.



\bibliographystyle{IEEEtran}
\bibliography{references}









\end{document}